\documentclass[10pt,conference]{IEEEtran}
\IEEEoverridecommandlockouts

\usepackage{cite}
\usepackage{amsmath,amssymb,amsfonts}
\usepackage{algorithmic}
\usepackage{graphicx}
\usepackage{textcomp}
\usepackage{xcolor}

\usepackage{graphicx}
\usepackage{textcomp}
\usepackage{amsmath}
\usepackage{listings}

\usepackage{tikz}
\usetikzlibrary{tikzmark}
\usepackage{subcaption}
\usepackage{xcolor}
\usepackage{capt-of}

\usepackage{multirow}

\lstdefinestyle{mystyle0}{
    numbers=left, 
    numberstyle=\tiny\color{gray}, 
    numbersep=5pt, 
    escapeinside={(*@}{@*)},
    frame=none,
    stepnumber=1
}

\def\BibTeX{{\rm B\kern-.05em{\sc i\kern-.025em b}\kern-.08em
    T\kern-.1667em\lower.7ex\hbox{E}\kern-.125emX}}
\begin{document}

\title{LightDE: A Lightweight Method for \\ Eliminating Dangling Pointers
}


\author{\IEEEauthorblockN{Xun An}
\IEEEauthorblockA{
    anxun@iie.ac.cn}

}


\maketitle

\begin{abstract}
    The widespread presence of Use-After-Free (UAF) vulnerabilities poses a serious threat to software security, 
    with dangling pointers being considered the primary cause of these vulnerabilities. 
    However, existing methods for defending against UAF vulnerabilities by eliminating dangling pointers need to interrupt the program's execution when encountering pointer assignment operations in order to store the memory addresses of the pointers in a specific data structure. 
    This makes these methods not lightweight. 
    To overcome this drawback, we propose a novel approach called LightDE. 
    This method does not require storing the memory addresses of pointers during program execution. 
    LightDE uses our proposed structure-sensitive pointer analysis method to determine which objects pointers point to and stores the pointing relationships in the program's data segment during program compilation. 
    Since LightDE only needs to verify if pointers identified by the pointer analysis point to released objects when eliminating dangling pointers, 
    it is very lightweight. 
    Our experimental results show that LightDE can effectively defend against UAF vulnerabilities and the performance overhead it introduces is very low. 
\end{abstract}

\begin{IEEEkeywords}
    Use-After-Free, Dangling Pointer, Pointer Analysis, LLVM
\end{IEEEkeywords}

\section{Introduction}\label{sec:intro}
\begin{figure}[htp]
\centering
\includegraphics[width=0.48\textwidth]{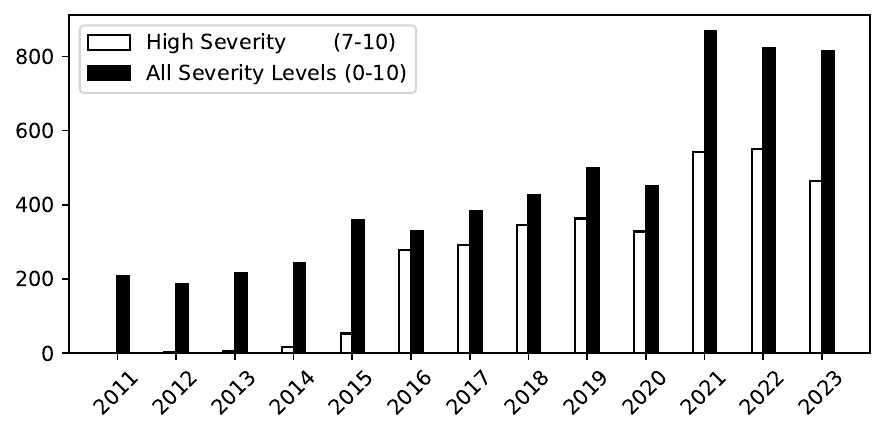}
\caption{Use-After-Free Vulnerabilities in NVD \cite{CVE-2020-16590334}.}
\label{fig:uafser}
\end{figure}

Use-After-Free (UAF) vulnerabilities pose a serious threat to software security as they can lead to arbitrary code execution, 
privilege escalation, and data leakage \cite{gui2021automated,song2019sok,szekeres2013sok}. 
Unfortunately, the number of UAF vulnerabilities has been increasing over the years, 
and their impact has become increasingly severe \cite{gui2021automated,yan2018spatio,cuoq2012frama,ye2014poster}. 
Table \ref{fig:uafser} shows the changes in the number of UAF vulnerabilities in the NVD database from 2011 to 2023, 
as well as the number of high-threat UAF vulnerabilities. 
There are two types of UAF vulnerabilities: stack UAF and heap UAF \cite{younan2015freesentry,van2018type,szekeres2013sok,an2023refining}. 
Stack UAF occurs when a dangling pointer is used to access stack memory that has already been freed. 
However, modern compilers generate warning messages for such vulnerabilities, making it easy for programmers to detect and fix them. 
Additionally, stack space allocation and deallocation are automatic, making stack UAF vulnerabilities difficult to exploit. 
Heap UAF refers to accessing freed heap memory through a dangling pointer. 
The difficulty in defending against heap UAF primarily stems from the flexible allocation and deallocation of heap memory, 
as well as the complexity of pointer propagation. 
All UAF vulnerabilities discussed in this paper are heap UAF. 

Researchers have proposed various methods to defend against UAF vulnerabilities, 
but each approach has its limitations and drawbacks. 
One method involves checking the validity of pointer access during pointer dereferencing, 
which requires verifying memory access legality before each pointer dereference \cite{nagarakatte2010cets,nagarakatte2012watchdog,necula2002ccured,condit2003ccured,zhang2019bogo}. 
However, this method can lead to incompatibility with existing libraries and introduce unacceptable runtime overhead. 
An alternative approach modifies memory allocation and deallocation strategies, 
making it harder for attackers to exploit UAF vulnerabilities \cite{boehm2002garbage,oiwa2009implementation,zhang2018use,novark2010dieharder,berger2006diehard,dang2017oscar}. 
However, this method only provides probabilistic defense, and attackers can bypass it through techniques like heap spraying and other attack methods. 
Recent research has proposed defending against UAF vulnerabilities by eliminating dangling pointers \cite{lee2015preventing,younan2015freesentry,van2017dangsan,liu2018robust,shen2020heapexpo}. 
Eliminating dangling pointers is a very promising method for mitigating UAF vulnerabilities, 
as dangling pointers are the root cause of these vulnerabilities. 
However, existing methods for eliminating dangling pointers need to record the memory addresses of pointers during program execution. 
This makes these methods not lightweight.

Since dangling pointers are the root cause of UAF vulnerabilities \cite{lee2015preventing,younan2015freesentry,van2017dangsan,liu2018robust,shen2020heapexpo}, 
this paper focuses on defending against UAF vulnerabilities by eliminating dangling pointers. 
Current methods \cite{lee2015preventing,younan2015freesentry,van2017dangsan,liu2018robust,shen2020heapexpo} for defending against UAF vulnerabilities by eliminating dangling pointers rely on inserting functions after pointer assignment instructions during compilation to track pointers. 
This method requires the program to pause its normal execution flow when encountering a pointer assignment instruction during runtime, and to execute the logic of the inserted function. 
The purpose of inserting functions is to record the memory addresses of pointers. 
Some methods \cite{lee2015preventing,younan2015freesentry,van2017dangsan,shen2020heapexpo} also need to locate the objects pointed to by the pointers. 
Although existing dangling pointer elimination schemes have designed various data structures to store the memory addresses of pointers to improve storage efficiency, 
their fundamental principles remain the same. 
In this process, 
recording the memory addresses of pointers and locating the objects they point to is time-consuming, 
which makes current dangling pointer elimination methods not lightweight.




To address the issue of existing methods being unable to eliminate dangling pointers in a lightweight manner, 
we propose LightDE, a lightweight approach to eliminate dangling pointers to defend against UAF vulnerabilities. 
LightDE is automatically deployed during program compilation and is completely transparent to the programmer. 
LightDE's uniqueness lies in its combination of our proposed structure-sensitive pointer analysis \cite{lhotak2013pointer,hardekopf2007ant,pereira2009wave,kroening2014cbmc} to determine which objects pointers reference at compile time, and storing these relationships in the program's data segment during compilation. 
The significant advantage of our method over existing methods is that it does not require storing the memory addresses of pointers or locating the objects pointed to by the pointers during program execution. 
When an object is released, 
LightDE checks whether the pointer identified by the pointer analysis still points to the released object based on the pointer relationships stored in the program's data segment, 
and eliminates the corresponding dangling pointers. 


The main challenge in implementing LightDE lies in identifying the objects pointed to by pointers at compile time. 
Field-sensitive pointer analysis methods can analyze the objects pointed to by pointer fields. 
However, field-sensitive pointer analysis results do not include the type information of the objects. 
When it is unclear which specific field of an object points to another object, 
field-sensitive analysis cannot infer which pointer fields within the object might point to that object. 
As a result, it cannot completely eliminate dangling pointers within the object. 
Structure-sensitive pointer analysis results include the type information of objects. 
This method can use the object's type information to infer the pointer fields within the object and their targets. 
But existing structure-sensitive pointer analysis cannot correctly analyze the case where an object has multiple types \cite{barbar2020flow,balatsouras2016structure} or cannot be used to analyze multi-threaded applications \cite{an2024structure}. 
To address this problem, we propose a structure-sensitive, flow-insensitive pointer analysis method based on Andersen's pointer analysis \cite{lhotak2013pointer,hardekopf2007ant,pereira2009wave,kroening2014cbmc} to acquire type information of objects. 
Our structure-sensitive pointer analysis is flow-insensitive, so our pointer analysis method can also be used to analyze pointer relationships in multithreaded applications. 
Although our pointer analysis method still has cases where it cannot identify the targets of pointer fields within objects, 
the results of our pointer analysis include the type information of objects. 
Therefore, when eliminating dangling pointers, we can check all the pointer fields of an object based on its type information to see if they point to the freed object. 
In Section \ref{sec:ssmotivation}, we will illustrate with an example that proposing a completely new structure-sensitive pointer analysis method is necessary for eliminating dangling pointers.




We implemented a prototype system of LightDE and demonstrated its effectiveness using real UAF vulnerabilities. 
Our experimental results on the SPEC CPU2006 \cite{henning2006spec} benchmark suite demonstrate that LightDE introduces very low runtime overhead, 
with a geometric mean runtime overhead of only 8.66\%. 
The geometric mean of the memory overhead introduced by LightDE is 7.26\%. 
Experimental results on the SPLASH-2X \cite{parsec2011memo} benchmark suite show that LightDE scales well on multi-threaded applications. 
pSweeper also runs a separate thread on idle CPU cores to eliminate dangling pointers, 
but the geometric mean of the runtime overhead introduced by pSweeper is 17.2\% which is significantly higher than the runtime overhead introduced by LightDE. 
Therefore, the reduced runtime overhead of LightDE is primarily due to determining and storing pointer relationships at compile time.

To summarize, our contributions in the field of defending against UAF vulnerabilities are as follows:

1. We propose a lightweight method for eliminating dangling pointers, LightDE. 
Our approach determines pointer points-to relationships during program compilation and embeds this information into the program's data segment at compile time. 
This allows our method to avoid interrupting the program's execution flow during runtime to store the memory addresses of pointers or locate pointer targets, 
making LightDE very lightweight. 


2. We propose a structure-sensitive pointer analysis method based on Andersen's pointer analysis to obtain object structure information and improve the accuracy of pointer analysis results. 


3. We developed a prototype system of LightDE based on LLVM-13 and SVF to validate our design. 
We conducted experiments to evaluate the LightDE prototype. 
The experimental results show that LightDE can effectively defend against UAF vulnerabilities while introducing very low performance overhead.



\begin{figure}[htbp]
    \centering
    \begin{minipage}{0.40\textwidth} 
        \centering
    \begin{lstlisting}[style=mystyle0]
    typedef struct {
        int* arr;
    } A;
    void main() {
        A* a = (A*)malloc(8);     (*@\color{green!60!black}//o1@*)
        (*@\color{red}storePtr(\&a);@*)
        int* b = (int*)malloc(4);  (*@\color{green!60!black}//o2@*)
        (*@\color{red}storePtr(\&b);@*)
        a->arr = b;
        (*@\color{red}storePtr(\&a--\textgreater arr);@*)
        ...
        free(b);
        ... = *a->arr;
    }
    
    void storePtr(void** ptrAddr) {
      (*@\color{green!60!black}// store the pointer address@*)
        ...
    }
    \end{lstlisting}
    \end{minipage}
    \caption{A Motivation Example.}
    \label{fig:exampleInsert}
    \end{figure}

    \begin{figure*}[htbp]
        \centering
        \includegraphics[width=1.0\textwidth]{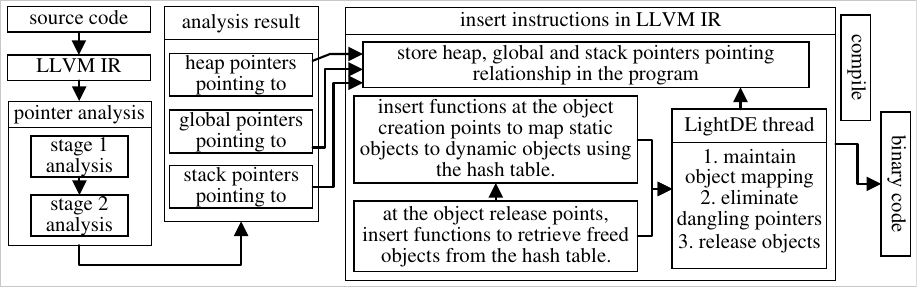}
        
        \caption{The Compile-Time Workflow of LightDE.}
        \label{fig:lightDEOverview}
        \end{figure*}
        
\section{Motivation Example}\label{sec:motivaExam}


To further illustrate the issue that existing methods of eliminating dangling pointers \cite{lee2015preventing,younan2015freesentry,van2017dangsan,liu2018robust,shen2020heapexpo} are not lightweight, 
we use the example in Figure \ref{fig:exampleInsert} to explain this problem. 
In this example, a type $A$ object $o1$ is created on line 5, 
and the memory address of $o1$ is stored in the pointer variable $A* \ a$. 
On line 7, an integer object $o2$ is allocated on the heap, 
and the memory address of $o2$ is stored in the pointer variable $int* \ b$. 
On line 9, the memory address of $o2$ is assigned to the first field of object $o1$. 
The code segment highlighted in red in Figure \ref{fig:exampleInsert} is an abstraction of the inserted instructions from the existing dangling pointer elimination methods. 
The function $storePtr$ in Figure \ref{fig:exampleInsert} is also abstracted from the existing dangling pointer elimination methods. 
The primary responsibility of this function is to record the pointer's address in the designated global data structure. 
It is worth noting that various methods may have slight differences in the specific operations of this function, 
but the basic functionality is similar. 
That is, tracking the pointers in order to eliminate the corresponding dangling pointers when memory is freed. 
When the program executes the code on line 6 of Figure \ref{fig:exampleInsert}, 
the $storePtr$ function records the memory address of $A* \ a$. 
Similarly, the code on lines 8 and 10 will trigger similar processes. 
This design means that whenever the program encounters a pointer assignment operation, 
it records the pointer's memory address. 
This process severely impacts the program's performance. 
Through the example in Figure \ref{fig:exampleInsert}, it is easy to understand why existing dangling pointer elimination methods are not lightweight.






\section{Overview of LightDE}\label{sec:overviewOf}



\begin{figure}[htbp]
\centering
\includegraphics[width=0.47\textwidth]{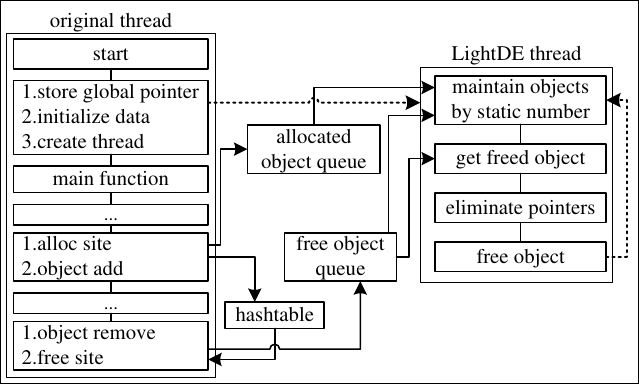}

\caption{The Runtime Workflow of LightDE.}
\label{fig:lightDETimeOverview}
\end{figure}

We use Figure \ref{fig:lightDEOverview} to illustrate the workflow and content insertion of LightDE during compilation, 
and use Figure \ref{fig:lightDETimeOverview} to illustrate the working state of LightDE during program execution. 
The workflow of LightDE at compile time can be divided into five parts: 
generating LLVM IR, performing pointer analysis, 
organizing the pointer analysis results, 
inserting the pointer analysis results and corresponding instructions into the LLVM IR, 
and recompiling the LLVM IR to generate the executable file. 
First, we compile the program's source code into LLVM Intermediate Representation (LLVM IR) using the LLVM compiler. 
LightDE performs pointer analysis based on the LLVM IR. 
Our structure-sensitive pointer analysis \cite{balatsouras2016structure,barbar2020flow,pearce2007efficient,avots2005improving} is conducted in two phases. 
The primary purpose of the first phase of pointer analysis is to obtain the structural information of objects. 
The second phase of pointer analysis uses the object structure information obtained in the first phase to filter out spurious pointing relationships. 
Next, based on the results of pointer analysis, LightDE categorizes pointers into three types based on the objects they point to: heap pointers, global pointers, and stack pointers. 
Based on the results of the pointer analysis and the pointer types, LightDE embeds the pointer pointing information in the program's data segment. 
However, it is not possible to determine the runtime memory address of a pointer in the static case. 
We will explain our method for storing pointer pointing relationships at compile time and mapping the results of static pointer analysis to runtime information in Section \ref{sec:static2Run}.


In the pointer analysis phase, 
our method of abstracting objects is to treat each heap allocation point as an object. 
This abstraction method is too coarse for eliminating dangling pointers, 
so we need to accurately identify every object that the program allocates and deallocates at runtime. 
To address this issue, 
LightDE uses a multi-threaded hash table \cite{gao2005lock} to map static objects to runtime objects. 
LightDE inserts a function after each object allocation point in the program to store the allocated object information and the static information of the object in our multi-threaded hash table. 
Similarly, LightDE inserts a function before each object release point to obtain the released objects from the hash table and pass their information to the LightDE thread. 
LightDE removes the program's original memory release function, 
allowing the LightDE thread to handle memory space release to prevent UAF vulnerabilities during the elimination of dangling pointers. 
The creation of the LightDE thread occurs when the program starts executing. 
This thread has two functions. 
The first function is to organize all runtime objects created by the same allocation point together after receiving the object information from the program’s original thread. 
This organization facilitates the lookup of corresponding runtime objects based on static object information when eliminating dangling pointers. 
The second function is to eliminate corresponding dangling pointers based on the pointing relationships stored in the program's data segment after receiving the information about released objects from the program’s original thread. 
Finally, the LLVM IR with the inserted data and functions is compiled into a binary program. 

Figure \ref{fig:lightDEOverview} only illustrates the workflow of LightDE at compile time, 
but it does not reflect the operation of LightDE at runtime. 
Therefore, we use Figure \ref{fig:lightDETimeOverview} to explain the operation of LightDE at runtime. 
When the program starts executing, 
LightDE initializes our data structures and creates the LightDE thread. 
When the program executes to the object allocation instruction, 
the next instruction after the object allocation is our inserted function for tracking runtime objects. 
The function's role is to store the starting memory address of the runtime object, 
the size of the object's memory, 
and related static object information into a global multithreaded hash table. 
Additionally, this function stores this information in a multithreaded lock-free queue. 
Our thread retrieves object information from this lock-free queue and organizes all runtime objects created at the same allocation point together. 
When the program executes to the deallocation point for the object, 
the original deallocation function of the program has already been removed by LightDE at compile time. 
Therefore, our object deallocation management function is executed instead. 
The role of our object deallocation management function is to remove the deallocated object's information from the global multithreaded hash table and store the deallocated object's information into a lock-free queue shared with the LightDE thread. 
The LightDE thread obtains the object information to be released from this queue, 
and check whether the relevant pointer points to the object to be released based on the pointing relationship pre-stored in the program data segment. 
In Figure \ref{fig:lightDETimeOverview}, we only show the entire process of allocating and releasing an object for one thread. 
However, the basic process for multiple objects and multiple threads is similar to the process shown in Figure \ref{fig:lightDETimeOverview}.

\section{Design}\label{sec:design}


In this section, 
we first explain LightDE's structure-sensitive pointer analysis method to obtain the pointing relationships of pointers and the structural information of objects. 
Then, we explain how LightDE maps static pointer analysis results to runtime information, enabling accurate elimination of dangling pointers.


\subsection{Structure-sensitive Pointer Analysis}\label{sec:pointerAna}

In this section, we first demonstrate through examples that our structure sensitivity is indispensable for eliminating dangling pointers. 
Then we explain our method for abstracting programs, 
and finally, we describe our pointer analysis method.

\subsubsection{Motivating Example For Our Structure-Sensitivity}\label{sec:ssmotivation}

\begin{figure}[htbp]
\centering
\begin{minipage}{0.45\textwidth} 
    \centering
\begin{lstlisting}[style=mystyle0]
typedef struct{
    void* pf;
} T1;
typedef struct{
    void *ps;
    void* px;
} T2;
void main() {
    int n;
    T1* p = (T1*)malloc(16);(*@\color{green!60!black}//o1@*) 
    ...
    T2* q = (T2*)p;
    q->px = malloc(8);      (*@\color{green!60!black}//o2@*)
    void* ptr = malloc(16); (*@\color{green!60!black}//o3@*)
    scanf("%d",&n);         (*@\color{green!60!black}//n = 0 or 8@*)
    void* p1 =(void*) p + n;  
    *((void**)p1) = ptr;  
    ...                        
    free(ptr);              (*@\color{green!60!black}//o3@*)
    ...
}
\end{lstlisting}
\end{minipage}
\caption{A Motivating Example for the Necessity of Our Structure-Sensitivity.}
\label{fig:exampleMYSS}
\end{figure}

We illustrate the innovation, necessity, and role of our structure-sensitive pointer analysis through the example in Figure \ref{fig:exampleMYSS}. 
In the code shown in Figure \ref{fig:exampleMYSS}, an object $o1$ is allocated on line 10 and cast to type $T1$, 
followed by a series of insignificant operations. 
On line 12, the code casts object $o1$ to type $T2$. 
On line 13, the code stores the memory address of object $o2$ into the $px$ field of object $o1$ through type $T2$. 
Existing structure-sensitive pointer analysis methods \cite{barbar2020flow,balatsouras2016structure} assume that an object has only one type throughout its entire lifecycle. 
Therefore, in this example, these existing methods assume that object $o1$ has only type $T1$. 
As a result, at line 13, they incorrectly conclude that object $o1$ does not have a $px$ field, 
leading to the inference that object $o1$ does not point to object $o2$. 
Additionally, the recent structure-sensitive pointer analysis method \cite{an2024structure} cannot be used to analyze multi-threaded applications. 
The code on line 16 in the figure \ref{fig:exampleMYSS} modifies the pointer $p$ using the integer $n$, and the pointer $p1$ may point to the first field of object $o1$ or the second field of object $o1$. 
Therefore, it is impossible to ascertain which field of object $o1$ points to object $o3$ from analyzing the code at line 17. 
Field-sensitive pointer analysis will infer that object $o1$ points to object $o3$.
However, since the field-sensitive pointer analysis results do not include the type information of the objects, it cannot infer the possible field offsets of object $o1$ that point to $o3$. 
This leads to an inability to eliminate the dangling pointer in object $o1$ after object $o3$ is freed on line 19. 
Similarly, our structure-sensitive pointer analysis results also cannot accurately determine which field of object $o1$ points to object $o3$. 
However, our structure-sensitive pointer analysis results include the type information $T1$ and $T2$ of object $o1$. 
Based on the type information of the object, our method can infer that either the field of object $o1$ at offset 0 points to object $o3$, 
or the field of object $o1$ at offset 8 points to object $o3$. 
After object $o3$ is released on line 19, 
LightDE can check the pointer fields at offsets 0 and 8 of object $o1$ to see if they point to object $o3$. 
The example in Figure \ref{fig:exampleMYSS} demonstrates that proposing a structure-sensitive pointer analysis method suitable for multithreading and multiple struct objects is essential for eliminating dangling pointers.

\subsubsection{Program Abstraction}\label{sec:absOfProgram}


This section includes three parts: 
abstraction of program domains, 
abstraction of program instructions, 
and symbols used in pointer analysis.

\begin{table}[htbp]
  \caption{Analysis Domains.}
\label{tab:domain}
\centering
\begin{tabular}{llll}
\hline
\multicolumn{4}{c}{Analysis domains}                                                                            \\ \hline
$t$                 & $\in$ & $\mathcal{T}$                                         & program types        \\
$fld$               & $\in$ &  $\mathcal{F}$                                        & fields             \\
$f$                 & $\in$ &  $\mathcal{C}$                                        &  field offsets                  \\
$s$                 & $\in$ & $\mathcal{S}$                                         & virtual registers         \\
$g$                 & $\in$ & $\mathcal{G}$                                         & global variables          \\
$m$                 & $\in$ & $\mathcal{M}$                                       & program functions         \\
$p,q,x,a$           & $\in$ & $\mathcal{P}$ $=$ $\mathcal{S}$ $\cup$  $\mathcal{G}$ & top-level variables       \\
$o$                 & $\in$ & $\mathcal{O}$                                         & abstract objects          \\
$\widehat{o_{f}}$ & $\in$ & $\mathcal{Y}$ $=$ $\mathcal{O \times F}$              & abstract field of objects \\
$d$                 & $\in$ & $\mathcal{A}$ $=$ $\mathcal{O}$ $\cup$  $\mathcal{Y}$ & address-taken variables   \\
$v$                 & $\in$ & $\mathcal{V}$ $=$ $\mathcal{P}$ $\cup$  $\mathcal{A}$ & program variables         \\ \hline
\end{tabular}
\end{table}


The domains related to our pointer analysis are shown in Table \ref{tab:domain}, 
mainly including variables, functions, types, and objects. 
Variables in programs $\mathcal{V}$ can be divided into two types: 
address-taken variables $\mathcal{A}$, which can be accessed through pointers, 
and top-level variables $\mathcal{P}$, which are never accessed through pointers. 
Address-taken variables $\mathcal{A}$ include abstract objects $\mathcal{O}$ and fields of abstract objects $\mathcal{Y}$. 
Heap objects are a type of address-taken object, 
and our method for abstracting heap objects is to treat each heap allocation point as a distinct object. 
We represent object fields by creating a sub-object for each field and establishing a relationship between the sub-object and its parent object. 
The creation of sub-objects is based on the offset of their fields relative to the base object. 



\begin{table}[htbp]
\caption{LLVM-like Instructions.}
\label{tab:inst}
\centering
\begin{tabular}{ll}
\hline
\multicolumn{2}{c}{LLVM-like instructions} \\ \hline
$p = alloca \ t,nbytes$    & STACK/GLOBAL  \\
$p = malloc \ nbytes$      & HEAP          \\
$p = q$                    & COPY          \\
$p = (t)q$                 & CAST          \\
$p = *q$                   & LOAD          \\
$*p = q$                   & STORE         \\
$p = \phi(q,h)$            & PHI           \\
$p = \&q \rightarrow fld$  & FIELD         \\
$p =q (...,a,...)$         & CALL          \\
$ret \ p$                  & RET           \\ \hline
\end{tabular}
\end{table}

\begin{figure*}[]
    \centering
    
    \begin{subfigure}{1.0\textwidth}
        \begin{align*}
        \begin{array}{l}
            \textbf{{\scriptsize [STACK/GLOBAL]}} \\
            $$ \ \   p = alloca \ t,nbytes$$   \\
            \hline
            $$\  \ pt(p) = pt(p) \cup \{o\}   \\  
            pt.t(o) = pt.t(o) \cup \{t\}$$
        \end{array} \ & \quad \ \ \ \  \ \
        \begin{array}{l}
            \textbf{{\scriptsize [HEAP]}} \\
            $$ \ \ p = malloc \ nbytes$$   \\
            \hline
            pt(p) = pt(p) \cup \{o\} $$ 
        \end{array} \ & \quad \ \  \ \
        \begin{array}{l}
            \textbf{{\scriptsize [COPY]}} \\
            $$ \ \ \ \ \ \ p = q    $$    \\
            \hline
            $$ pt(p) \supseteq pt(q)  $$ 
        \end{array} \ & \quad \ \ \  \ \
        \begin{array}{l}
            \textbf{{\scriptsize [CAST]}} \\
            $$ p = (t) q \ \ \   \ \  o    \in    pt(q) $$   \\
            \hline
            $$ \ \ \ \ \  pt(p) \supseteq pt(q)    \\  
            pt.t(o) = pt.t(o) \cup \{t\}$$
        \end{array}
        \end{align*}
    \end{subfigure}  
    
    
    \begin{subfigure}{1.0\textwidth}
        \begin{align*}
        \begin{array}{l}
            \textbf{{\scriptsize [PHI]}} \\
            $$ \ \ \ \ \ \  p = \phi(q,x)   $$     \\
            \hline
            $$   pt(p) \supseteq pt(q) \cup  pt(x) $$ 
        \end{array} \ & \quad \ \  \ \
        \begin{array}{l}
            \textbf{{\scriptsize [LOAD]}} \\
            $$  p = *q \ \   o  \in  pt(q)$$   \\
            \hline
            $$ \ \  pt(p) \supseteq pt(o)  $$ 
        \end{array}  \ & \quad \ \  \
        \begin{array}{l}
            \textbf{{\scriptsize [STORE]}} \\
            $$  *p = q \ \  o    \in    pt(p)      $$   \\
            \hline
            $$\ \    pt(o) \supseteq  pt(q)  $$ 
        \end{array} \ & \quad \ \ \quad \  \ \
        \begin{array}{l}
            \textbf{{\scriptsize [FIELD-VAR]}} \\
            $$  p = \&q \rightarrow *   \  \ \  o    \in    pt(q)  $$ \\
            \hline
            $$\ \  pt(p) = pt(p) \cup  \{ \widehat{o_{\ast}} \}  $$ 
        \end{array}
        \end{align*}
    \end{subfigure}

    \begin{subfigure}{1.0\textwidth}
        \begin{align*}
        \begin{array}{l}
            \textbf{{\scriptsize [FIELD]}} \\
            $$  p = \&q \rightarrow fld    \ \ f = off(q \rightarrow fld)  \ \  o  \in  pt(q) $$ \\
            \hline
            $$ \ \ \ \ \ \ \ \ \ \ \ \  pt(p) = pt(p) \cup \{\widehat{o_{f}}\}  $$  
        \end{array} \ & \  \
        \begin{array}{l}
            \textbf{{\scriptsize [CALL]}} \\
            $$ p = q(...,a,...)  \ \   m(...,a',...)  \ \   m \in pt(q)    $$   \\
            \hline
            $$ \ \ \ \ \ \ \  \ \ \ \ \   pt(a')  \supseteq pt(a) $$ \\
        \end{array} \ & 
        \begin{array}{l}
            \textbf{{\scriptsize [RET]}} \\
            $$  p = q(...) \ \       ret \ p'    $$  \\
            \hline
            $$ \ \  pt(p)  \supseteq pt(p') $$
        \end{array}
        \end{align*}
        
    \end{subfigure}

    \vspace{-0.1cm}

    \caption{Stage 1 Pointer Analysis Rules.}
    \label{fig:AnderWithType}
    
    \end{figure*}

Our pointer analysis is based on the LLVM IR of programs. 
LLVM defines a rich set of instructions, 
but we can abstract these instructions into a small instruction set to represent all program instructions. 
Our abstracted instruction set is shown in Table \ref{tab:inst}. 
STACK/GLOBAL instructions allocate stack or global objects. 
HEAP instructions are used for heap object allocation. 
The COPY instruction is an assignment statement. 
PHI is the convergence point of a branching structure that selects different values based on the branch taken. 
CAST is used to interpret an object as a specific type to satisfy LLVM's type checking without altering the contents in memory. 
LOAD and STORE are operations for reading from or writing to address-taken objects. 
FIELD obtains the address of an object's field. 
CALL and RET instructions are used for function calls and returns.


The symbols used in this paper for pointer analysis are as follows: 
$pt(p)$ represents the set of objects pointed to by pointer $p$. 
For example, $pt(p) = \{o1, o2\}$ means that pointer $p$ points to objects $o1$ and $o2$. 
$pt.t(o)$ represents the set of types owned by object $o$. 
$pt.o(t)$ represents the set of field offsets that type $t$ owns. 
The function $off()$ is used to obtain the offset of a field within an object. 
The symbol $\widehat{o_{f}}$ denotes the sub-object of object $o$ with an offset of $f$. 
The symbol $*$ represents any field offset of an object. 
The symbol $\supseteq$ denotes a containment relationship, meaning the elements of the right set are added to the left set. 
For example, $pt(p) \supseteq pt(q)$ means the elements of the set $pt(q)$ are added to the set $pt(p)$. 
The rest of the symbols have the same meaning as the mathematical set operators.


\subsubsection{Pointer Analysis}\label{sec:inferenceRule}


The instructions for pointer analysis can be categorized into five types: 
object creation, 
object propagation, 
type conversion, 
field access, 
and function call/return instructions. 
The inference rules for the first phase of pointer analysis are shown in Figure \ref{fig:AnderWithType}.

Object creation involves two rules: [STACK/GLOBAL] and [HEAP]. 
[STACK/GLOBAL] is used to create stack and global objects, 
and it can obtain the type information of objects from the creation instruction. 
In this rule, an object $o$ is created, 
and we use $pt(p) = pt(p) \cup \{o\}$ to indicate that the pointer $p$ points to a new object $o$. 
The type of object $o$ in this rule is $t$, 
and we use $pt.t(o) = pt.t(o) \cup \{t\}$ to indicate that the object acquires type $t$. 
The [HEAP] rule only involves the creation of heap objects.


Object propagation involves the four rules [COPY], [PHI], [LOAD], and [STORE]. 
These rules only involve the propagation of objects. 
Both the [COPY] and [PHI] rules are simple assignment instructions, 
propagating the points-to information of the right-hand pointer to the left-hand pointer. 
Specifically, the [PHI] rule merges the points-to sets of the two pointers on the right and propagates them to the left-hand pointer. 
[LOAD] and [STORE] propagate objects with the help of a third pointer. 


Type conversion only involves the [CAST] rule. 
Here, we briefly explain our type model. 
To ensure the accuracy and completeness of object type information, 
an object's type set is only allowed to grow and not shrink. 
We adopt this method because various types of pointers pointing to an object do not disappear when the object's type changes. 
These pointers can still access fields of the object's previous types. 
Thus, we cannot discard an object's existing types when its type changes. 
In the [CAST] rule, the target type is added to the object's type set. 

Field access instructions involve the [FIELD] and [FIELD-VAR] rules. 
[FIELD] is used to analyze cases where an object's field is accessed by a known offset. 
In such cases, the pointer points to the offset of the corresponding field of the object. 
[FIELD-VAR] is used to analyze cases where an object's field is accessed by an unknown offset. 
In these cases, we conservatively assume that the pointer might point to any field of the object. 
Additionally, it is worth noting that we treat arrays as a special kind of structure. 
The elements within an array have the same type, 
but each element has a different offset. 
Therefore, our method is also array-sensitive.

Function calls involve the [CALL] and [RET] rules. 
We employ an On-The-Fly approach for resolving function calls. 
This method further analyzes the call relationships between functions based on the results of pointer analysis. 
This kind of function call analysis can effectively analyze indirect function calls, such as $call$ $*q$. 


\begin{figure}[htbp]
    \centering

    \begin{align*}
    \begin{array}{l}
        \textbf{{\scriptsize [FIELD]}} \\
        $$\ \ \ \ \ \ \ \ p = \&q \rightarrow fld \ \ \ \ \ \ \ \ \ \ f = off(q \rightarrow fld )  $$ \\
        \hline
            $$ \forall  o_i \in pt(q) \land t \in pt.t(o_i) \land f \in pt.o(t)  ,  pt(p) \supseteq \{\widehat{o_{if}}\}   $$  \\
        \end{array}
    \end{align*}
    \vspace{-0.1cm}
    
    \caption{Stage 2 Pointer Analysis Rule.}
    \label{fig:andersenUseType}
    
    \end{figure}

The second phase of pointer analysis uses the object type information obtained in the first phase to filter out spurious pointers pointing to object fields. 
The inference rules in the second phase are similar to those in the first phase, 
with only the [CAST] and [FIELD] rules being different. 
Since the type information of objects has already been obtained in the first phase, 
the [CAST] rule is not needed in the second phase. 
Type conversion instructions are treated as copy instructions. 
The [FIELD] rule in the second phase checks whether each type in the object's type set contains the field offset being accessed by the current statement. 
If the field offset cannot be found, the points-to relationship is filtered out. 
If the field offset exists, a sub-object is created for that offset, and the pointer points to the created sub-object. 
The [FIELD] rule is shown in Figure \ref{fig:andersenUseType}.

The specific pointer analysis process involves the following steps: 
During stage 1 of pointer analysis, 
we repeatedly apply the pointer analysis rules in Figure \ref{fig:AnderWithType} until the pointer pointing set no longer changes, 
marking the end of stage 1. 
Then, stage 2 of pointer analysis begins. 
This stage is similar to stage 1 but uses the object type information obtained from the first stage to make the pointer analysis more accurate. 
The pointer analysis concludes when the pointer pointing set no longer changes. 

We conducted a soundness analysis of our method. 
When analyzing instructions that access object fields, 
our method uses the type information of the objects to check whether the object has a field at the corresponding offset, 
making the pointer analysis results more accurate. 
However, there may be cases where the entire object is accessed as raw memory without using structural information. 
For objects that do not contain structural information, 
we do not use structural information to filter their points-to relationships. 
Our method is capable of accurately analyzing both structured and unstructured objects, thereby ensuring its soundness.

\subsection{Mapping Static Results to Runtime}\label{sec:static2Run}

In this section, we explain the method of mapping static pointer analysis results to runtime information of the program. 
First, we describe the method of mapping static objects to runtime objects. 
Then, we explain the method of mapping static pointers to runtime pointers. 
Finally, we explain our method for accurately eliminating dangling pointers by utilizing the pointer analysis results and the program's runtime information. 

\subsubsection{Object Mapping}\label{sec:ObjMap}

There are two reasons why we need to map static objects to runtime objects: 
1. A single static object corresponds to multiple runtime objects. 
2. The memory addresses of objects can only be determined at runtime. 
During pointer analysis, 
we abstract each heap allocation point as a specific object. 
This abstraction method allows us to quickly analyze points-to relationships during pointer analysis, 
but it is not precise enough for eliminating dangling pointers. 
This is because the same allocation point can create multiple different objects based on different contexts. 
Therefore, it is unreasonable to continue using this abstraction method when eliminating dangling pointers. 


To distinguish different objects created at the same allocation point, 
we use the runtime information of the program. 
We insert a function after each allocation point in the program. 
The main purpose of this inserted function is to store the allocated memory information in our designed multi-threaded hash table and to pass the information of the created objects to our thread through a multi-threaded lock-free queue \cite{valois1995lock,aldinucci2012efficient}. 
The LightDE thread groups all dynamic objects created at the same allocation point together, 
enabling us to find the corresponding runtime objects based on the static object. 
The object information stored in our data structure includes the memory start address, end address, size, 
and the identifier of the corresponding static object.



\subsubsection{Pointer Mapping}\label{sec:ptrMap}

In this section, we explain our method for mapping static pointers to runtime pointers. 




\textbf{Heap Pointer Mapping} 
Before introducing our method for mapping heap pointers, 
we use an example to illustrate the relationship between static heap pointers and runtime heap pointers. 
Suppose there are two objects, $o1$ and $o2$. 
The $f$ field of $o2$ points to the object $o1$. 
In static analysis, $o1$ and $o2$ correspond to two allocation points, 
which may correspond to many runtime objects. 
Therefore, $o2.f$ corresponds to multiple heap pointers at runtime. 
Our mapping method involves allocating a data structure for each static object based on the pointer analysis results during program compilation. 
The data structure records the heap pointers pointing to the object. 
Since the addresses of heap pointers cannot be determined during compilation, 
we record the static identifier of the static object containing the heap pointer and the offset of the heap pointer within the static object in the data structure. 
When an object is released, 
our method for eliminating heap pointers that point to the released object is to check the heap pointers at the same offset location in the runtime objects corresponding to the static object that contains the heap pointers. 
If we cannot determine the exact offset of the heap pointer within the object, 
we check all the object's pointer fields to see if they point to the freed object. 
Since our pointer analysis is structure-sensitive, 
we can determine all the offsets of the object's pointer fields.

\textbf{Global Pointer Mapping} 
In this paragraph, 
we explain our method for mapping static global pointers to runtime global pointers. 
Global pointers can be accessed from any location in the program, 
but their memory addresses cannot be determined during program compilation. 
To address this issue, 
we assign indices based on the names of global pointers. 
During compilation, 
we store the indices of global pointers that point to a certain static object in its pointer set. 
We insert a function at the beginning of the program's startup function, 
which stores the memory addresses of global pointers into a global array according to their indices. 
Using this global array, we can check whether the global pointers that point to a static object are pointing to the corresponding released runtime object.


\textbf{Stack Pointer Mapping} 
In this paragraph, 
we explain our method for mapping static stack pointers to runtime stack pointers. 
Our method involves assigning a static identifier to each function of the program during compilation and assigning a static index to all stack pointers within each function in the order they are defined. 
During compilation, we store the function identifier that the stack pointer resides in and the stack pointer's static index within the function into the object's pointer set based on the pointer analysis results. 
In each function, 
we insert instructions to store the stack pointer into our data structure based on the static index within the function. 
When an object is released, we check whether the stack pointers identified by the pointer analysis point to the released object. 
Here, we can also see the shortcomings of our method for handling stack pointers. 
Our method still requires storing the memory addresses of stack pointers at runtime, but it avoids looking up the objects pointed to by the pointers. 
Since we use a separate thread to asynchronously eliminate stack pointers, 
we need to avoid the stack space being freed while eliminating stack pointers. 
We designed a dedicated stack for this purpose. 
When the program needs to allocate stack pointers, 
it allocates space on this dedicated stack. 
However, the original thread of the program does not release the allocated space. 
When a function exits, the program's original thread generates exit information and passes it to our thread. 
Our thread then releases the corresponding stack space. 

\subsubsection{Dangling Pointer Elimination}\label{sec:danglingPtrEli}

In this paragraph, 
we explain the method LightDE uses to eliminate corresponding dangling pointers after an object is released. 
We insert our memory management function before the release points identified by the pointer analysis and remove the program's original memory release function. 
The LightDE thread releases the memory after setting the pointers pointing to the released memory to invalid pointers to prevent UAF vulnerabilities during the elimination of dangling pointers. 
Our memory management function removes the information of the released object from our multi-threaded hash table and passes this information to the LightDE thread. 
The LightDE thread then retrieves the static identifier of this runtime object and uses it to obtain the static pointers stored in the program's data segment that point to the object. 
Next, the LightDE thread uses the pointer mapping method described in \ref{sec:ptrMap} to check whether the corresponding runtime pointers point to the released object, 
and it invalidates the corresponding dangling pointers.

\begin{table}[]
    \caption{Comparison the Accuracy of Our Structure-sensitivity with Andersen's Pointer Analysis.}
    \label{tab:accuracy}
\centering
\begin{tabular}{llll}
\hline
Benchmark      & \begin{tabular}[c]{@{}l@{}}Andersen\\  Points-to\#\end{tabular} & \begin{tabular}[c]{@{}l@{}}Our method\\   Points-to\#\end{tabular} & \begin{tabular}[c]{@{}l@{}}Points-to\#\\  Filtered out\end{tabular} \\ \hline
400.perlbench  & 82704268                                                             & 82195642                                                               & 508626                                                              \\
401.bzip2      & 23845                                                                & 23097                                                                  & 748                                                                 \\
403.gcc        & 516822633                                                            & 512698520                                                              & 4124113                                                             \\
429.mcf        & 2293                                                                 & 2071                                                                   & 222                                                                 \\
433.milc       & 30626                                                                & 30454                                                                  & 172                                                                 \\
444.namd       & 21422                                                                & 20347                                                                  & 1075                                                                \\
445.gobmk      & 518615                                                               & 518588                                                                 & 27                                                                  \\
447.dealII     & 9557789                                                              & 9033265                                                                & 524524                                                              \\
450.soplex     & 12375851                                                             & 11314009                                                               & 1061842                                                             \\
453.povray     & 45313149                                                             & 39046254                                                               & 6266895                                                             \\
456.hmmer      & 218420                                                               & 208138                                                                 & 10282                                                               \\
458.sjeng      & 12398                                                                & 12281                                                                  & 117                                                                 \\
462.libquantum & 2975                                                                 & 2948                                                                   & 27                                                                  \\
464.h264ref    & 1484659                                                              & 1464771                                                                & 19888                                                               \\
470.lbm        & 789                                                                  & 789                                                                    & 0                                                                   \\
471.omnetpp    & 154075153                                                            & 71838897                                                               & 82236256                                                            \\
473.astar      & 5025                                                                 & 5008                                                                   & 17                                                                  \\
482.sphinx3    & 227546                                                               & 209173                                                                 & 18373                                                               \\
483.xalancbmk  & 126216408                                                            & 119172228                                                              & 7044180                                                             \\ \hline
\end{tabular}
\end{table}

\section{Evaluation}\label{sec:experiment}

Our experiments first evaluate the accuracy improvement of our structure-sensitive pointer analysis compared to Andersen's pointer analysis. 
Then, our experiments assess three aspects of LightDE: effectiveness, performance overhead, and scalability on multi-threaded applications. 
First, we experimentally assess LightDE's effectiveness in defending against real UAF vulnerabilities. 
Next, we evaluate the performance overhead introduced by LightDE using the SPEC CPU2006 benchmark suite \cite{henning2006spec}. 
Finally, we assess LightDE's scalability on multi-threaded applications using the SPLASH-2X benchmark suite \cite{parsec2011memo}. 
We believe that heap pointers and global pointers exist for a longer duration and have a wider scope of influence, 
making them more likely to cause UAF vulnerabilities. 
Therefore, in our experiments, 
we analyzed the impact on program performance when considering all pointers, including stack, heap, and global pointers, 
as well as the impact when considering only heap and global pointers. 
In the figure of the experimental results, 
the label "with-stack" represents the case where all pointers are considered, 
while the label "no-stack" represents the case where only heap and global pointers are considered. 
We implemented a prototype system of LightDE based on LLVM-13 and SVF \cite{sui2016svf,sui2014detecting,sui2012static,svf@website}. 
All experiments were conducted on a machine running 64-bit Ubuntu 20.04, equipped with a 13th Gen Intel(R) Core(TM) i7-13700H processor at 2.640GHz and 32GB of RAM.

\subsection{Accuracy of our Sturcture-sensitivity}\label{sec:accuracy}


Before assessing the performance of LightDE in eliminating dangling pointers, 
we first evaluate the accuracy improvement of our structure-sensitive pointer analysis method compared to Andersen pointer analysis on the SPEC CPU2006 benchmark suite. 
Our structure-sensitive pointer analysis method is a two-phase pointer analysis. 
The first phase of pointer analysis is similar to Andersen's pointer analysis. 
The difference between the first stage and the Andersen's is that our method simultaneously determines the type information of objects while analyzing pointer targets. 
The second stage of pointer analysis filters out false pointer relationships by using the object type information obtained in the first stage. 
The main objective of this experiment is to evaluate how many spurious pointing relationships our structure-sensitive pointer analysis reduces compared to Andersen pointer analysis. 
The first and second columns of Table \ref{tab:accuracy} show the number of pointer relationships inferred by Andersen's pointer analysis and the number of pointer relationships inferred by our structure-sensitive pointer analysis for each item, respectively. 
The third column of the table indicates the number of spurious point-to relationships reduced by our structure-sensitive pointer analysis compared to Andersen pointer analysis for each benchmark. 
From the data in the last column of the table, we can see that the number of spurious point-to relationships filtered out by our structure-sensitive pointer analysis varies significantly across different benchmarks. 
For instance, in benchmark 470.lbm, our method shows no accuracy improvement over Andersen pointer analysis, 
whereas in benchmark 471.omnetpp, our method filtered out more than 50\% of the false point-to relationships present in Andersen pointer analysis.
The primary reason for this discrepancy lies in the nature of the programs themselves. 
Our structure-sensitive pointer analysis method filters false point-to relationships based on the types of objects, 
so our method requires knowledge of the exact offsets used by the instructions to access the objects when filtering false point-to relationships.
If the program accesses object fields through variable offsets, 
our method will be unable to filter out false point-to relationships based on object types. 
These variable offsets may require complex analysis or be influenced by input data, as illustrated by the example in Figure \ref{fig:exampleMYSS}. 
Therefore, if there are a large number of cases in the program where object fields are accessed via unknown offsets, 
our method will be unable to filter out false pointing relationships based on object types.

\begin{table}[htbp]
    \caption{Real-world UAF Vulnerabilities.}
    \label{tab:cveTest}
    \centering
  \begin{tabular}{llc}
  \hline
  CVE ID         & Application         & Protected                 \\ \hline
  CVE-2024-24260 \cite{CVE-2024-24260} & media-server v1.0.0 & \checkmark \\
  CVE-2020-16590 \cite{CVE-2020-16590} & Binutils 2.35       & \checkmark \\
  CVE-2018-5747 \cite{CVE-2018-5747} & lrzip 0.631         & \checkmark \\
  CVE-2017-12448 \cite{CVE-2017-12448} & Binutils 2.29       & \checkmark \\ \hline
  \end{tabular}
  \end{table}

  \begin{table*}[htbp]
    \caption{Statistics for SPEC CPU2006.}
    \label{tab:static}
    \centering
    \begin{tabular}{ll@{\hspace{1.5em}}l@{\hspace{1.5em}}lll|l@{\hspace{1.5em}}l@{\hspace{1.5em}}ll}
    \hline
    Benchmark      & \begin{tabular}[c]{@{}l@{}}static \\ obj \#\end{tabular} & \begin{tabular}[c]{@{}l@{}}free \\ site \#\end{tabular} & \begin{tabular}[c]{@{}l@{}}static heap\\ pointer \#\end{tabular} & \begin{tabular}[c]{@{}l@{}}static global\\ pointer \#\end{tabular} & \begin{tabular}[c]{@{}l@{}}static stack\\ pointer \#\end{tabular} & \begin{tabular}[c]{@{}l@{}}runtime\\ alloc obj \#\end{tabular} & \begin{tabular}[c]{@{}l@{}}runtime \\ free obj \#\end{tabular} & \begin{tabular}[c]{@{}l@{}}invalidate\\ pointer \#\end{tabular} & \begin{tabular}[c]{@{}l@{}}pSweeper \\ invalidate\\ pointer \#\end{tabular} \\ \hline
    400.perlbench  & 22                                                          & 253                                                     & 357                                                              & 3626                                                               & 2033                                                              & 358M                                                           & 356M                                                           & 18M        & 421K                                                           \\
    401.bzip2      & 6                                                           & 7                                                       & 18                                                               & 3                                                                  & 0                                                                 & 162                                                            & 144                                                            & 18          & 0                                                     \\
    403.gcc        & 1724                                                        & 1201                                                    & 2021962                                                          & 281490                                                             & 759291                                                            & 21M                                                            & 21M                                                            & 2938K        & 186K                                                    \\
    429.mcf        & 4                                                           & 9                                                       & 27                                                               & 10                                                                 & 0                                                                 & 3                                                              & 3                                                              & 4            & 0                                                    \\
    433.milc       & 61                                                          & 30                                                      & 416                                                              & 37                                                                 & 18                                                                & 6509                                                           & 6468                                                           & 0            & 0                                                    \\
    444.namd       & 44                                                          & 402                                                     & 21                                                               & 6                                                                  & 41                                                                & 1329                                                           & 1321                                                           & 56            & 0                                                   \\
    445.gobmk      & 51                                                          & 9                                                       & 298                                                              & 65                                                                 & 238                                                               & 656K                                                           & 656K                                                           & 209K           & 86                                                   \\
    447.dealII     & 388                                                         & 1502                                                    & 28384                                                            & 12347                                                              & 49348                                                             & 24M                                                            & 24M                                                            & 14M             & 7293                                                  \\
    450.soplex     & 269                                                         & 469                                                     & 52972                                                            & 13609                                                              & 3145                                                              & 67K                                                            & 67K                                                            & 473             & 553                                                 \\
    453.povray     & 532                                                         & 583                                                     & 284014                                                           & 45194                                                              & 302064                                                            & 2460K                                                          & 2460K                                                          & 1747K           & 7428                                                  \\
    456.hmmer      & 224                                                         & 330                                                     & 11754                                                            & 1                                                                  & 538                                                               & 1351K                                                          & 1351K                                                          & 709K            & 0                                                 \\
    458.sjeng      & 10                                                          & 16                                                      & 27                                                               & 7                                                                  & 0                                                                 & 4                                                              & 0                                                              & 0               & 0                                                 \\
    462.libquantum & 7                                                           & 7                                                       & 0                                                                & 2                                                                  & 5                                                                 & 95                                                             & 95                                                             & 0               & 0                                                  \\
    464.h264ref    & 293                                                         & 582                                                     & 7446                                                             & 429                                                                & 4                                                                 & 176K                                                           & 176K                                                           & 828K            & 961                                                 \\
    470.lbm        & 2                                                           & 2                                                       & 0                                                                & 4                                                                  & 0                                                                 & 2                                                              & 2                                                              & 1               & 0                                                 \\
    471.omnetpp    & 2434                                                        & 2155                                                    & 831479                                                           & 204579                                                             & 49198                                                             & 312M                                                           & 266M                                                               & 4694k        & 13K                                                          \\
    473.astar      & 64                                                          & 81                                                      & 138                                                              & 47                                                                 & 17                                                                & 4790K                                                          & 4790K                                                          & 3049K            & 104                                                \\
    482.sphinx3    & 162                                                         & 93                                                      & 5225                                                             & 100                                                                & 9                                                                 & 7918K                                                          & 7918K                                                          & 4115K            & 762                                                \\
    483.xalancbmk  & 564                                                         & 5605                                                    & 253507                                                           & 192478                                                             & 443941                                                            & 126M                                                           & 126M                                                           & 18M              & 79K                                                \\ \hline
    \end{tabular}
    \end{table*}

\subsection{Effectiveness}\label{sec:effectiveness}

To evaluate the effectiveness of LightDE in mitigating UAF vulnerabilities, we applied LightDE to four programs with known UAF vulnerabilities. 
We then ran exploitation programs to determine LightDE's effectiveness in defending against these vulnerabilities. 
The experimental results show that LightDE was able to prevent the exploitation of UAF vulnerabilities in all four applications, as shown in Table \ref{tab:cveTest}.

\subsection{Performance on SPEC CPU2006}

In this section, we evaluate the performance overhead of LightDE on the SPEC CPU2006 benchmark suite.

\subsubsection{Statistics}

Table \ref{tab:static} shows the static data of the SPEC CPU2006 benchmark suite. 
The second column displays the number of static objects for each project. 
The seventh column shows the number of objects allocated at runtime. 
By comparing these two columns, 
it is evident that a single static object corresponds to multiple runtime objects. 
The third column shows the number of release points identified by the pointer analysis, 
while the eighth column shows the number of objects released at runtime. 
From these two columns, it is also clear that a single release point releases multiple runtime objects. 
The fourth column shows the number of static heap pointers identified by the pointer analysis, 
the fifth column shows the number of static global pointers identified by the pointer analysis, 
and the sixth column shows the number of static stack pointers identified by the pointer analysis. 
The ninth column displays the number of dangling pointers eliminated by LightDE. 
The last column of the table shows the number of dangling pointers eliminated by pSweeper \cite{liu2018robust} in each project. 
Based on the last two columns of the table, it can be observed that LightDE eliminates significantly more dangling pointers than pSweeper.



\begin{figure}[htbp]
    \centering
    \includegraphics[width=0.48\textwidth]{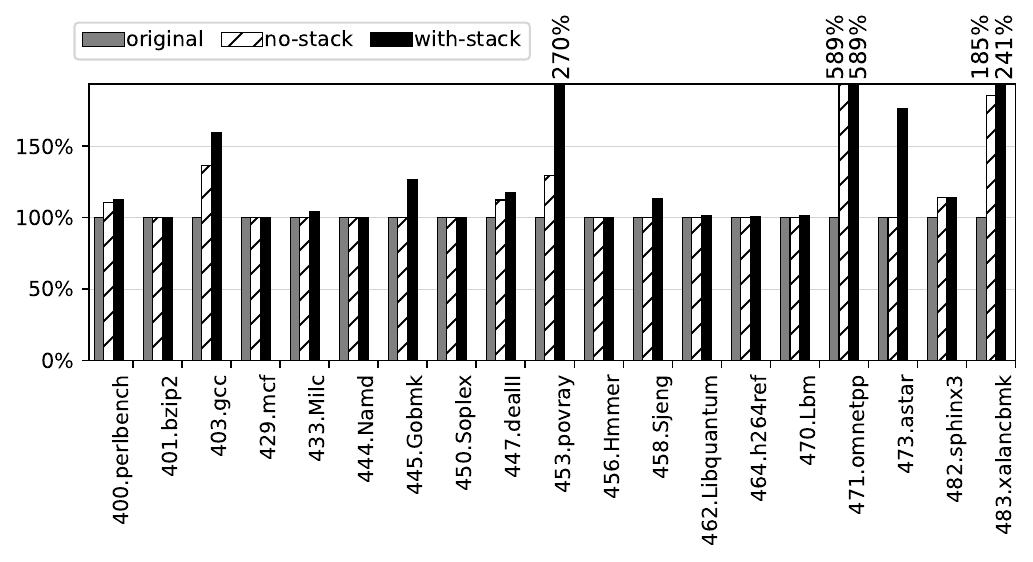}
    \caption{LightDE's Runtime Overhead on SPEC CPU2006.}
    \label{fig:sepcTime}
    \end{figure}

    \begin{figure*}[htbp]
        \centering
        \includegraphics[width=1.0\textwidth]{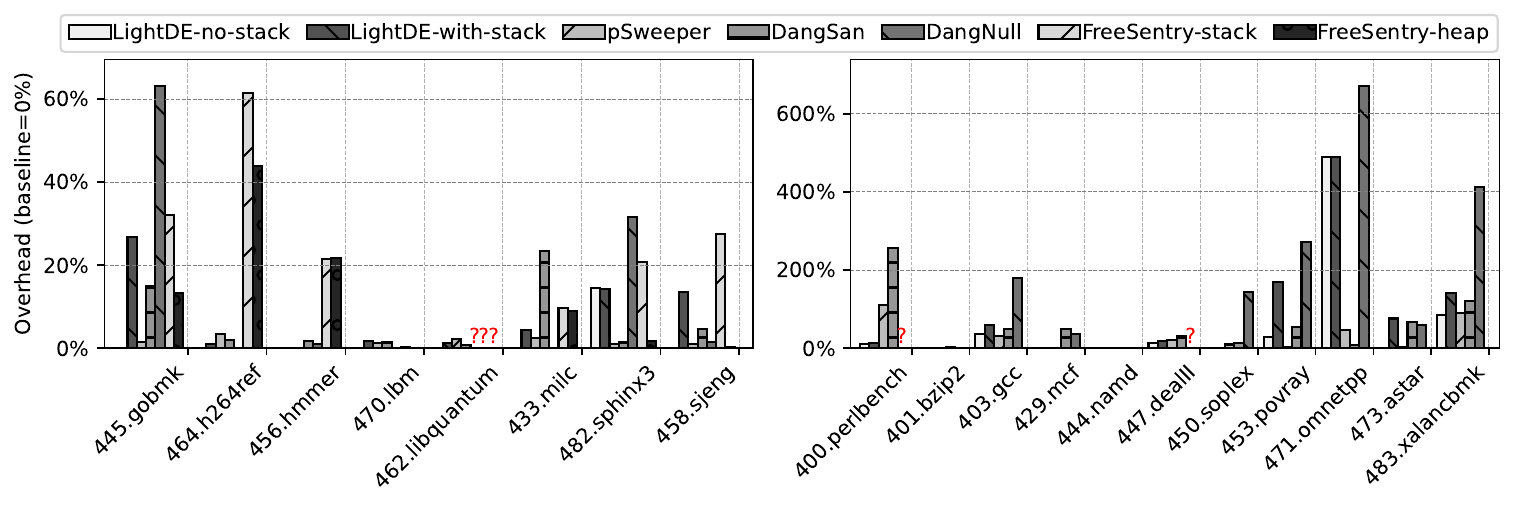}
        \caption{Runtime Overhead Comparison. '?' indicates the corresponding overhead was not reported in the method.}
        \label{fig:alltimcmp}
        \end{figure*}

    \begin{figure}[htbp]
        \centering
        \includegraphics[width=0.48\textwidth]{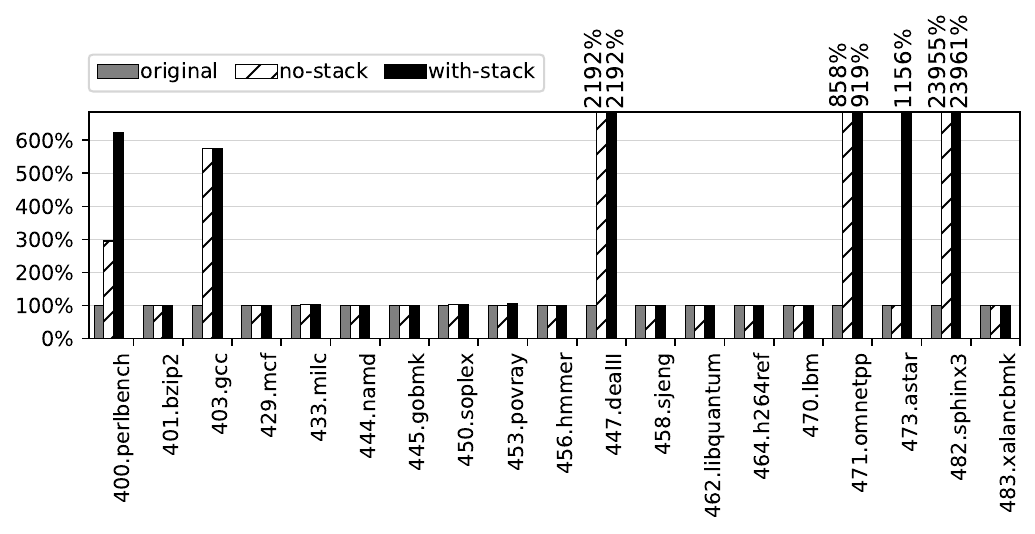}
        \caption{LightDE's Memory Overhead on SPEC CPU2006.}
        \label{fig:cpuMem}
        \end{figure}
    
\subsubsection{Runtime Overhead}


Figure \ref{fig:sepcTime} shows the runtime overhead introduced by LightDE in each project. 
When only the heap and global pointers are considered, 
LightDE introduces an additional geometric mean runtime overhead of 3.76\% compared to the unprotected original program. 
When considering all pointers, including heap, stack, and global pointers, 
LightDE introduces an additional geometric mean runtime overhead of 8.66\% compared to the unprotected original program. 
It is worth noting that the time or memory overhead introduced by LightDE in some projects is far below 0.01\%. 
To calculate the geometric mean, we set the overhead for these projects to 1\% in this paper.






As shown in Figure \ref{fig:sepcTime}, when LightDE considers only heap and global pointers, 
the runtime overhead introduced in most projects is very low. 
However, we also observed significant runtime overhead in the 471.omnetpp and 483.xalancbmk projects. 
The reason for this is that these projects are memory allocation-intensive applications. 
These two projects frequently allocate and release memory, 
requiring frequent maintenance of the hash table and the lock-free queue, 
which leads to significant performance overhead. 
As shown in Figure \ref{fig:sepcTime}, when LightDE considers all pointers, including stack, heap, and global pointers, the runtime overhead increases. 
This is because our method requires the original thread of the program to store the memory addresses of stack pointers in our designed data structure and notify our thread to remove the corresponding local pointers upon function exit. 
For example, in the 453.povray and 473.astar projects, the frequent function entry and exit caused relatively high runtime overhead. 






\subsubsection{Memory Overhead}


Figure \ref{fig:cpuMem} shows the memory overhead introduced by LightDE in each project. 
When considering only heap and global pointers, 
LightDE introduces a geometric mean memory overhead of 3.88\% compared to the unprotected original program. 
When considering all pointers, including heap, stack, and global dangling pointers, 
LightDE introduces a geometric mean memory overhead of 7.26\%.

As shown in Figure \ref{fig:cpuMem}, when considering only heap and global pointers, 
the memory overhead introduced by LightDE is very low in most projects. 
However, in the projects 447.dealII, 471.omnetpp, and 482.sphinx3, LightDE introduced significantly high memory overhead. 
We analyzed the reasons for the high memory overhead introduced by LightDE in these projects. 
The memory overhead of LightDE mainly comes from the inability to timely release objects that are awaiting release. 
Since the results of pointer analysis can be very inaccurate under certain circumstances, 
it can lead to LightDE thread checking too many irrelevant pointers. 
This ineffective checking leads to excessive time consumption, 
causing the objects awaiting release to accumulate in the shared queue between the original thread and the LightDE thread, 
resulting in higher memory overhead. 
When considering stack pointers, 
we observed that in the 473.astar project, 
a significantly high memory overhead was introduced. 
This is due to the frequent function calls in the 473.astar project, 
which generated a large number of stack pointers, 
resulting in very high memory consumption. 



\begin{figure*}[htbp]
\centering
\includegraphics[width=1.0\textwidth]{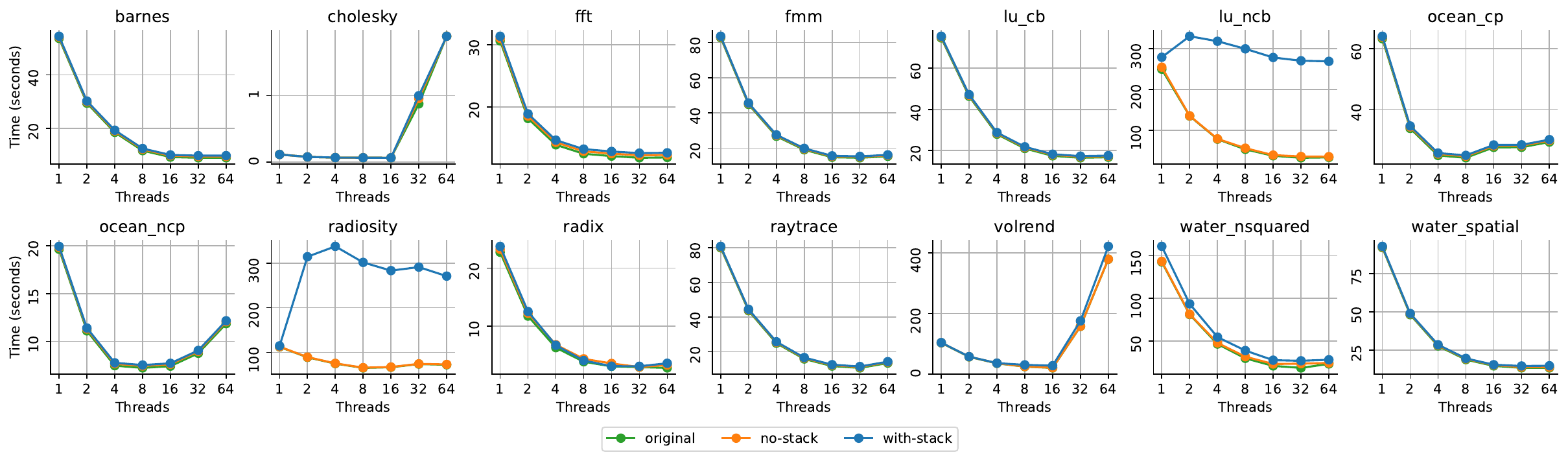}
\caption{LightDE's Runtime Overhead on SPLASH-2X.}
\label{fig:splash2XTime}
\end{figure*}

\begin{figure*}[htbp]
\centering
\includegraphics[width=1.0\textwidth]{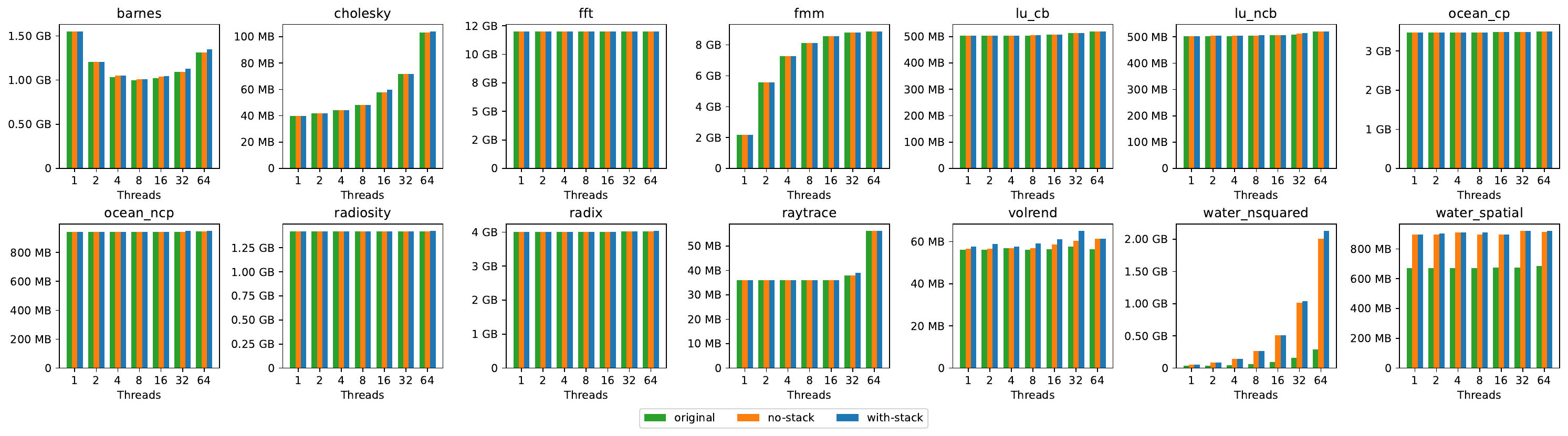}
\caption{LightDE's Memory Overhead on SPLASH-2X.}
\label{fig:splash2XMem}
\end{figure*}

\subsubsection{Overhead Comparison}


In this section, 
we compare the runtime and memory overhead of LightDE with existing methods of eliminating dangling pointers \cite{lee2015preventing,younan2015freesentry,van2017dangsan,liu2018robust,shen2020heapexpo}. 
The runtime overhead introduced by existing methods and LightDE is shown in Figure \ref{fig:alltimcmp}. 
Since FreeSentry only reports results of 7 SPEC CPU2006 projects, 
we display all the projects reported by FreeSentry in the left image of Figure \ref{fig:alltimcmp}. 

The geometric mean of the runtime overhead introduced by DangNull in its reported projects is 55\%. 
It is worth noting that DangNull only considers heap pointers and does not take stack pointers and global pointers into account. 
FreeSentry reported the overhead of eliminating only heap dangling pointers and the overhead of eliminating both heap and stack pointers. 
Since FreeSentry reported too few SPEC CPU2006 projects, 
we do not calculate the geometric mean of its runtime overhead. 
The geometric mean of the runtime overhead introduced by DangSan is 41\%. 
The geometric mean of the runtime overhead introduced by pSweeper is 17.2\%. 
When considering only global and heap pointers, 
the geometric mean of the runtime overhead introduced by our method is 3.76\%. 
When considering all pointers, the geometric mean of the runtime overhead is 8.66\%. 
Our method introduces significantly lower runtime overhead compared to existing dangling pointer elimination methods, 
whether considering stack pointers or not. 

We only compared memory overhead with the better performing dangling pointer elimination methods DangSan and pSweeper. 
The geometric mean of the memory overhead introduced by DangSan is 240\%, 
and the geometric mean of the memory overhead introduced by pSweeper is 247.3\%. 
The memory overhead introduced by LightDE is significantly lower than that of existing dangling pointer elimination methods, 
regardless of whether stack pointers are considered.

During our experiments, we had a question: 
Is LightDE's low runtime overhead mainly due to the fact that it determines and stores pointer relationships during compilation, or is it due to multithreading? 
By comparing the specific processes of eliminating dangling pointers in pSweeper and our method, LightDE, we can reaches the conclusion that the reduced runtime overhead of LightDE is primarily due to determining and storing pointer relationships at compile time. 
pSweeper requires the original thread to record the information of allocated objects and the memory addresses of pointers during program execution, 
and the original thread records the released object information into a specific data structure when an object is released. 
And pSweeper uses a separate thread to scan the pointers recorded in the global data structure to check whether they point to released objects. 
LightDE similarly requires the original thread to record the information of allocated objects during program execution, 
and when an object is released, the original thread stores the information of the released object into a specific data structure. 
LightDE similarly uses a separate thread to check whether the pointers identified by static pointer analysis still point to the released object. 
Since the separate threads that pSweeper and LightDE run on idle CPUs do not affect the program's runtime overhead, 
the analysis of runtime overhead mainly compares the operations of pSweeper and LightDE on the original threads. 
The only difference between pSweeper and LightDE on the original thread is that pSweeper requires the original thread to record the memory addresses of pointers during program execution. 
Therefore, we can conclude that LightDE's reduced runtime overhead is mainly due to determining and storing the pointer relationships at compile time. 


\subsection{Scalability on Multi-threaded Applications}

We used the SPLASH-2X benchmark suite to evaluate LightDE's scalability on multi-threaded applications. 
Figures \ref{fig:splash2XTime} and \ref{fig:splash2XMem} show the memory and runtime overhead introduced by LightDE compared to the original program. 
We ran all the projects in the SPLASH-2X benchmark suite. 
We evaluated the overhead of setting each program to run 1, 2, 4, 8, 16, 32, and 64 threads. 
We evaluated the overhead of using LightDE to eliminate only heap and global dangling pointers, 
as well as the overhead of using LightDE to eliminate all dangling pointers, including heap, stack, and global dangling pointers.

Figure \ref{fig:splash2XTime} shows the runtime overhead introduced by LightDE on the SPLASH-2X benchmark suite. 
When considering only heap and global pointers, 
the geometric mean of the runtime overhead caused by LightDE is 1.23\%. 
When considering all pointers, including heap, stack, and global dangling pointers, 
the geometric mean of the runtime overhead is 3.58\%. 
From Figure \ref{fig:splash2XTime}, it can be seen that when only heap and global pointers are considered, 
LightDE introduces almost no runtime overhead. 
In this case, the runtime of the programs protected by LightDE overlaps with the runtime of the unprotected programs. 
However, when LightDE considers stack, heap, and global dangling pointers, 
it results in significant runtime overhead in the "lu\_ncb" and "radiosity" projects. 
We analyzed this phenomenon. 
In our design, the memory addresses of stack pointers are stored in our data structure by the original program threads. 
This indicates that these two projects experience frequent function entries and exits, 
causing the original program threads to frequently store the memory addresses of stack pointers. 



Figure \ref{fig:splash2XMem} shows the memory overhead introduced by LightDE on the SPLASH-2X benchmark suite. 
When considering only global and heap pointers, 
the geometric mean of the memory overhead introduced by LightDE is 2.10\%. 
When considering all pointers, including heap, stack, and global pointers, 
the geometric mean of the memory overhead is 2.31\%. 
From Figure \ref{fig:splash2XMem}, it can be seen that in most projects, 
the memory overhead introduced by LightDE is negligible, 
regardless of whether stack pointers are considered. 
However, the projects "water\_nsquared" and "water\_spatial" result in relatively higher memory overhead. 
In the "water\_nsquared" project, memory overhead increases with the number of threads, 
indicating that a large number of objects are created by the program's child threads rather than by the main thread. 
In the "water\_spatial" project, 
the memory increase remains almost constant across all thread counts, 
suggesting that the increase in memory overhead does not correlate with the number of threads. 
This implies that the memory overhead primarily comes from the main thread, which creates more objects. 





\section{Limitations and Future Work}\label{sec:limitation}

We have demonstrated through experiments that LightDE is very effective at eliminating dangling pointers. 
However, LightDE still has a major limitation. 
The limitation of LightDE arises from the imprecision of static object abstraction. 
This inaccuracy is due to the fact that the object abstraction method we use in the pointer analysis is too coarse. 
In our pointer analysis, we abstract the memory allocation point as a single object. 
In a static case, when a pointer at offset $f$ of static object $o1$ points to another static object $o2$, 
the corresponding runtime case is that all runtime objects created at the allocation point of $o1$ are considered to have pointers at offset $f$ that point to all runtime objects created at the allocation point of $o2$. 
This leads to checking too many irrelevant pointers when eliminating heap dangling pointers. 
This excessive checking of irrelevant pointers consumes a lot of time during the elimination process, 
causing the LightDE thread to be unable to timely release objects that are pending release, 
which are passed through the lock-free queue by the original program's thread, 
resulting in very high memory overhead. 
In our future work, 
we will investigate more accurate object abstraction methods.

\section{RELATED WORK}\label{sec:relate}

Existing methods to defend against UAF vulnerabilities by eliminating dangling pointers include DangNULL \cite{lee2015preventing}, FreeSentry \cite{younan2015freesentry}, DangSan \cite{van2017dangsan}, HeapExpo \cite{shen2020heapexpo}, pSweeper \cite{liu2018robust}, and ISDE \cite{an2023refining}. 
DangNULL and FreeSentry insert a specific function after pointer assignment instructions during compilation. 
This function is used to look up the objects pointed to by the pointers and add the pointers to the object's pointer set. 
DangNULL and FreeSentry maintain the objects pointed to by the pointers in real time. 
When the object pointed to by a pointer changes, 
DangNULL and FreeSentry remove the pointer from the pointer set of the previously pointed-to object and add it to the pointer set of the newly pointed-to object. 
DangNULL uses a tree structure to store the object's pointer set, while FreeSentry uses a two-level mapping table. 
DangSan uses a method similar to a system log to record pointers. 
DangSan inserts a function after pointer assignment instructions to add the pointer to the pointer set of the object it points to. 
If the target of a pointer changes, it does not remove the pointer from the set of the previously pointed-to object. 
This approach avoids the real-time maintenance of object pointer sets, 
resulting in lower performance overhead. 
When an object is freed, DangSan checks whether the pointers in the object's pointer set still point to the object. 
HeapExpo uses the same method as DangSan but prohibits the compiler from optimizing local pointers into registers, 
resulting in very high performance overhead. 
To further improve performance, 
pSweeper adopts an asynchronous method to eliminate dangling pointers. 
During program execution, pSweeper only stores pointers in its carefully designed global data structure, 
avoiding the need to look up the objects pointed to by the pointers. 
pSweeper runs a separate thread on an idle CPU core. 
This thread continuously scans the pointers recorded in the global data structure to check if they point to freed memory and eliminates dangling pointers. 
However, all of these methods require recording the memory addresses of pointers at runtime, and DangNULL, FreeSentry, DangSan, and HeapExpo also need to locate the objects pointed to by the pointers during program execution. 
Finally, it is worth noting that the previous method, ISDE, also attempted to eliminate dangling pointers by combining pointer analysis techniques \cite{an2023refining}. 
However, ISDE mainly focused on eliminating dangling pointers in registers. 
ISDE uses a flow-sensitive pointer analysis method to determine the targets of pointers. 
However, the flow-sensitive pointer analysis method can only analyze the objects pointed to by pointers in single-threaded applications. 
Therefore, ISDE can only eliminate dangling pointers in single-threaded programs.

\section{Conclusion}

This paper proposes LightDE, 
a lightweight method for eliminating dangling pointers to defend against UAF vulnerabilities. 
LightDE stores pointer points-to information in the program's data segment at compile time based on our pointer analysis results, 
avoiding the need to store pointer memory addresses and look up pointer targets during program execution. 
To achieve this, we present a structure-sensitive pointer analysis method for analyzing the points-to information and the type information of objects. 
We implemented a prototype system of LightDE and validated its effectiveness and efficiency. 


\bibliographystyle{IEEEtran}
\bibliography{IEEEabrv,ref.bib}



\end{document}